\documentclass[aps,pra,floatfix,showpacs,noshowkeys,epsfig,graphics,natbib]{revtex4}%
\usepackage{graphicx}
\usepackage{amsmath}
\usepackage{amsfonts}
\usepackage{amssymb}

\newcommand{\newc}{\newcommand}
\newc{\beq}    {\begin{equation}}
\newc{\eeq}    {\end{equation}}

\newc{\beqa}    {\begin{eqnarray}}
\newc{\eeqa}    {\end{eqnarray}}
\newc{\bs}    {\section}
\newc{\no}    {\\ \nonumber}

\newc{\st}    {\stackrel}

\begin{document}
\title{Dark energy from vacuum entanglement}
\author{Jae-Weon Lee}\email{scikid@kias.re.kr}
\affiliation{School of Computational Sciences,
             Korea Institute for Advanced Study,
             207-43 Cheongnyangni 2-dong, Dongdaemun-gu, Seoul 130-012, Korea}

\author{Jungjai Lee}
\affiliation{Department of Physics, Daejin University, Pocheon, Gyeonggi 487-711, Korea}

\author{Hyeong-Chan Kim}
\affiliation{Department of Physics, Yonsei University,
Seoul 120-749, Republic of Korea.
}%
\date{\today}
\begin{abstract}
We suggest that  vacuum entanglement energy associated with the entanglement
entropy  of the universe is the origin of  dark energy.
 The observed properties of
  dark energy
can be explained by using the nature of
 entanglement energy  without modification of gravity or exotic matter.
From the number of  degrees of  freedom
in the standard model, we obtain the equation of state parameter $\omega^0_\Lambda\simeq -0.93$
and $d\simeq 0.95$ for the holographic dark energy, which are consistent with  current observational
 data at the $95\%$ confidence level.
\end{abstract}

\maketitle

The cosmological constant problem is one of the most important unsolved puzzles in modern
physics~~\cite{CC}.
There is strong evidence from  Type Ia supernova (SN Ia) observations~~\citep{riess-1998-116} that
the  universe is expanding at an accelerating rate.
A simple explanation  for this acceleration is the existence of negative pressure fluids, called
the dark energy, whose pressure $p_\Lambda$ and density $\rho_\Lambda$ satisfy
$\omega_\Lambda\equiv p_\Lambda/\rho_\Lambda<-1/3$. (See Eq. (\ref{friedmann})).
 Although, there are various dark energy models rely on materials such as
quintessence~\cite{PhysRevLett.80.1582},
$k$-essence~~\cite{PhysRevLett.85.4438}, phantom~~\cite{phantom}, and Chaplygin gas~~\cite{Chaplygin}
among many,
the identity of this dark energy remains a mystery.
These models usually require fine tuning of potentials or unnatural characteristics of the materials.
On the other hand,
entanglement (a nonlocal quantum correlation)~\cite{nielsen}
 is now treated as an important physical quantity.
The possibility of exploiting entanglement in
quantum
information processing applications such as quantum key distribution and
quantum teleportation
 has led to intense study of this quantity  by the quantum information community.
Recently, there has been  renewed
interest~\cite{ryu:181602,emparan-2006-0606,solodukhin:201601} in
studying  black hole entropy using entanglement
entropy~\cite{PhysRevD.34.373,hckim} in the context of the AdS/CFT
correspondence~\cite{Maldacena}.
 In this paper, we suggest that there is an unexpected
 relation between dark energy and entanglement
 which are the two most puzzling entities in modern physics.

It is well known~\cite{0264-9381-22-22-014} that a simple combination of the Planck scale
and IR cutoff $L$ ( of order of inverse of the Hubble parameter $H$) gives an energy density comparable to the observed cosmological
constant or  dark energy.
This can be understood in terms of the holographic principle
proposed by 't Hooft and Susskind~\cite{hooft-1993},
which is a  conjecture  claiming
 that all of  the information in a volume  can be described by the physics
  at the boundary
 of the volume  and that the maximum entropy in a volume is proportional to its surface area.
 Cohen et al~\cite{PhysRevLett.82.4971} proposed  a relation between
the UV cut-off $l$ of an effective theory and $L$ by
 considering that
the total energy in a region of size $L$ can not be larger than
the mass of a black hole of that size. Thus, for $L=H^{-1}$, the zero-point vacuum energy density is bounded as
 \beq
 \label{holodark}
\rho_\Lambda=l^{-4}\st{<}{\sim} \frac{M_P^2}{ L^2 } = M_P^2 H^2.
 \eeq
 Interestingly, saturating the bound gives $\rho_\Lambda$
comparable to the observed  dark energy density $\sim
10^{-10} eV^4$ for $H=H_0\sim 10^{-33}~eV$, the present Hubble
parameter. The   success of this estimation
over the naive estimate $\rho_\Lambda=O(M_P^4)$
can be attributed to the fact that quantum field theory  over-counts the
 independent physical degrees
of freedom inside the  volume.  Thus, dark energy models
based on the holographic principle have an  advantage over
other models in that they do not need an  $ad~hoc$ mechanism to
cancel the $O(M_P^4)$ zero-point energy of the vacuum.  This
simple holographic dark energy model
 is suggestive, but not without problems of its own.
 Hsu~\cite{hsu} pointed out that for $L=H^{-1}$,  the Friedmann
equation $\rho=3M^2_P H^2$ makes the dark energy behave like
matter rather than a negative
pressure fluid, and prohibits accelerating expansion of the
universe. Later, Li~\cite{1475-7516-2004-08-013} suggested that
 holographic dark energy of the form \beq \label{holodark2}
\rho_\Lambda=\frac{3 d^2 M_P^2}{ R_h^2 }, \eeq
 would  give an accelerating universe, where   the future event horizon ($R_h$) is used instead
 of  the Hubble horizon
 as the IR cutoff $L$.
  Here
$d$ is an $O(1)$ constant.
 However this use of $R_h$ has yet to be
 adequately justified.
Attempts~\cite{horvat:087301,0295-5075-71-5-712,gong:064029,pavon-2005-628}
have been made
  to overcome this IR cutoff problem in other ways,  for example,
  by using non-minimal coupling to a scalar field
~\cite{0295-5075-71-5-712,gong:064029} or interaction between dark energy
and dark matter ~\cite{pavon-2005-628}.
Despite  some success, the holographic dark energy models  usually lack
either   an explanation for the microscopic  origin of the  dark energy
or an explanation
for why $d$, the constant that determines the
characteristics of the
  dark energy, is approximately one.

In this paper we propose that  these problems can be  overcome in a natural manner
by identifying  dark energy as  entanglement energy associated with
the entanglement entropy $S_{Ent}$ of the universe. Our model also suggests a way to derive $d$ and $\omega_\Lambda$
from the standard model of  particle physics.

From the Reeh-Schlieder theorem~\cite{reeh}
it is  known that
 the vacuum
for  general quantum fields violates
 Bell inequality and has entanglement~\cite{werner,narnhofer1}
when there are causally disconnected regions.
Entanglements of  Bose~\cite{my} and Fermi~\cite{oh} states
have been studied using a thermal Green's function approach.
In Ref. ~\cite{our} it was suggested that the Hadamard Green's function
representing quantum fluctuation of the vacuum
is  useful
for the study of entanglement  in a scalar field vacuum.
These relations between  vacuum quantum fluctuations and entanglement
are reminiscent of
the vacuum fluctuation model of dark energy~\cite{padmanabhan-2005-22}.

There are two natural physics related to the event horizon; black hole physics and entanglement physics.
However, identifying $R_h$ as a black hole horizon is problematic, because dark energy should not
include ordinary matter energy, while black hole energy includes all the energy inside the horizon.
In quantum information theory, the event horizon plays a role of an information barrier and this leads
to modification of energy of subsystem inside the horizon, which is the entanglement energy.
Therefore, the vacuum entanglement energy is a  remaining plausible candidate for holographic dark energy.
The entanglement entropy is
the von Neumann entropy $S_{Ent}=-Tr(\rho_A log \rho_A)$ associated with the reduced density matrix
$\rho_A\equiv Tr_B \rho_{AB}$ of a bipartite system $AB$ described by a density matrix $\rho_{AB}$~\cite{nielsen}.
 For pure states such as the quantum fields
vacuum, $S_{Ent}$ is a good measure of entanglement. When there is
an event horizon,  a natural choice is to divide the system into
two subsystems - inside and outside  the event horizon - and to
trace over one of these subsystems to calculate the entanglement,
 because
the event horizon represents the global causal
structure~\cite{li-2004-603}. Thus, $S_{Ent}$ is intrinsically
related to the event horizon rather than the particle horizon or
the Hubble horizon.
 The future event horizon is given by
 \beq
\label{Rh} R_h\equiv R(t)\int_t^\infty \frac{d R(t')}{H(t')
R(t')^2},
\eeq
which can be used as a typical length scale of the
system with the horizon. Here we consider the flat ($k=0$)
Friedmann universe which is favored by observations~\cite{wmap} and
inflationary theory~\cite{inflation} and described by the metric
\beq
ds^2=-dt^2+R^2(t)d\Omega^2,
 \eeq
 where $R(t)$ is the scale factor
as usual.
The entanglement entropy of the quantum field vacuum with a horizon is
generally expressed in the form
\beq \label{Sent} S_{Ent}=\frac{\beta
R_h^2}{a^2},
\eeq
where $\beta$ is an $O(1)$ constant that
depends on the nature of the field.
Here, $a$ is the UV cut-off of quantum gravity and
different from $l$ which is the UV cut-off of a low energy effective theory~\cite{PhysRevLett.82.4971}
(see below for details).
 $S_{Ent}$ has a form
consistent with the holographic principle,  although it is derived
from quantum field theory without using the principle.
Entanglement entropy for a single massless scalar field in the
Friedmann universe is calculated in Ref.~\cite{PhysRevD.52.4512,PhysRevLett.71.666}.
By performing numerical calculations on a sphere lattice, they
obtained $\beta=0.30$. If there are $N_{dof}$  spin degrees of
freedom  of  quantum fields  in  $R_h$, due to the additivity
of the entanglement entropy~\cite{nielsen}, we can add up
the contributions from all of the individual fields to
$S_{Ent}$~\cite{PhysRevD.52.4512}, that is, $S_{Ent}=N_{dof} \beta
R_h^2/a^2$,
where  for simplicity we assume the
same $\beta$ for all fields.

In ~\cite{entenergy} the entanglement energy  $E_{Ent}$ is defined as disturbed vacuum energy due to the presence
of a boundary. There,
entanglement energy proportional to the radius of the spherical volume  was derived
 from quantum field theory.
Thus, for the event horizon, the entanglement energy is generally given by
\beq
\label{alpha}
E_{Ent}=\alpha R_h,
\eeq
where $\alpha$ is a constant depending on the exact mathematical definition of $E_{Ent}$.
We suggest that
this entanglement energy
is the origin of dark energy.
Once we obtain $\rho_\Lambda$ from $E_{Ent}$,
the negative pressure $p_\Lambda$ can be  derived from the conservation of energy momentum tensor,
\beq
\label{p}
p_\Lambda=\frac{d(R^3\rho_\Lambda)}{dR (-3 R^2)}
\eeq
as usually done in holographic dark energy models (see Eq. (6) of Ref. ~\cite{1475-7516-2004-08-013}).
Recall that this equation can be  derived from the Freedmann equation with perfect fluid
having a  energy momentum tensor of the form
\beq
\label{Tperfect}
T_{\mu\nu}=(\rho_\Lambda+p_\Lambda) U_\mu U_\nu- p_\Lambda g_{\mu\nu},
\eeq
where $U^{\mu}U_{\mu}=1$.
Eq. (\ref{p}) indicates that perfect fluid with increasing energy as the universe expands
has a negative pressure.
Then, it is straight forward to obtain $\omega_\Lambda=p_\Lambda/\rho_\Lambda$ (see Eq. (\ref{omega})).
We will show below that our theory gives the desired form of holographic dark energy. Thus, we can use
the all known formalism of typical holographic dark energy models for our model.

Now let us determine the coefficient $\alpha$ in Eq. (\ref{alpha}).
Although the mathematical definition of entanglement energy is not well-established, there are several
reasonable conjectures for $E_{Ent}$ in Ref.~~\cite{entenergy,Myung2}.
Inspired by the holographic principle, we adopt the following definition among them:
\beq
\label{dE}
dE_{Ent}\equiv  T_{Ent} dS_{Ent}.
\eeq
Note that this is $not$  the first
law of thermodynamics for $E_{Ent}$ which needs a pressure term $but$
a mere definition of $E_{Ent}$ we choose in this paper.
In ~\cite{entenergy}, it was shown that this definition for $E_{Ent}$ is good for black holes.
Our entanglement
energy in Eq. (\ref{dE})
is this modified vacuum energy   and hence ``internal" energy which looks like some ``thermal
energy" related to entanglement entropy.
To calculate $E_{Ent}$, the most natural choice for  the  ``temperature" related to
the event horizon is  the Gibbons-Hawking temperature
$T_{Ent}= 1/(2\pi R_h)$~\cite{PhysRevD.15.2738,izquierdo-2006-633,1475-7516-2006-09-011}.
 By integrating $dE_{Ent}$ we obtain
 \beq \label{eent}
E_{Ent}=\frac{\beta N_{dof}R_h}{\pi a^2}.
\eeq
 From Eq. (\ref{alpha}) and Eq. (\ref{eent}), we see $\alpha={\beta N_{dof}}/{\pi a^2}$.  Then, the entanglement
energy density within the event horizon is given by \beq
\label{rho}
\rho_{\Lambda}=\frac{3 E_{Ent}}{4 \pi R_h^3}=\frac{3 \beta
N_{dof} }{4 \pi^2 a^2 R_h^2}\equiv \frac{3 d^2 M_P^2  }{ R_h^2 },
\eeq which  has  the form (Eq. (\ref{holodark2})) for the
holographic dark energy.  From the above
equation  we  immediately obtain a  formula for the constant
 \beq
\label{d1}
 d=\frac{\sqrt{\beta N_{dof}}}{2\pi a M_P}
 \eeq
for the first time.
Although the constant $d$ determines the characteristic of the dark energy and
 the final fate of the universe,
it has been constrained only by  observations so far.

Interestingly, our model can be easily  verified by current observations.
  The equation of state for  dark energy of the form in Eq. (\ref{holodark})
   is as follows  ~\cite{li-2004-603,1475-7516-2004-08-006}
\beq
\label{omega}
\omega_{\Lambda} =-\frac{1}{3} \left(1+\frac{2\sqrt{\Omega_\Lambda} }{d}\right),
\eeq
where  $\Omega_\Lambda$ is the density parameter of the dark
energy.
Now, by inserting the expression for  $d$ in Eq. (\ref{d1})  into the above equation,
 we obtain $\omega_\Lambda$  directly
from the number of spin degrees of freedom $N_{dof}$ in the standard model(SM):
\beq
\label{omega2}
\omega_{\Lambda} =-\frac{1}{3} \left (1+\frac{4\pi a M_P\sqrt{\Omega_\Lambda} }{\sqrt{\beta N_{dof}}}\right).
\eeq
Since $aM_P\simeq 1$, $\beta\simeq 1$, and $N_{dof}= O(10^2)$,  Eqs. (\ref{d1}) and (\ref{omega}) gives us $d\simeq 1$ and
$\omega_\Lambda\simeq -1$.
Thus, the above calculation explains why $d\simeq 1$ from a particle physics view point.
More precisely, we choose natural values $a=1/M_P$, $\beta= 0.3$, the dark energy density parameter for the present
$\Omega_\Lambda=0.73$  and
the matter density parameter $\Omega_M= 0.27$  favored by recent observations~\cite{wmap3}.
Using  $N_{dof}=118$ for the SM,
  we obtain
$d\simeq 0.95$ and $\omega^0_\Lambda\simeq -0.93$ for the present.
 \begin{figure}[htbp]
\includegraphics[width=0.5\textwidth]{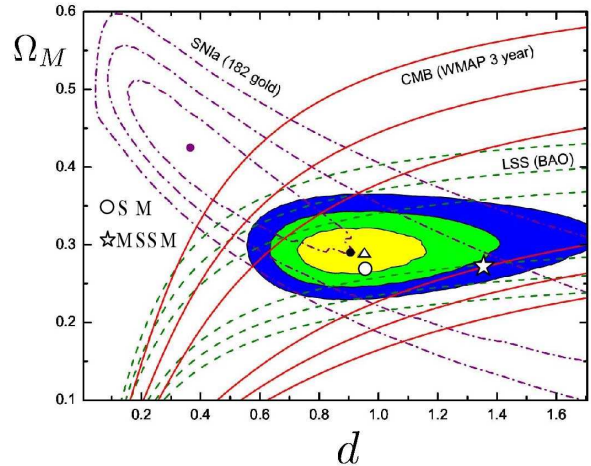}
\caption{Allowed parameter region for $d$  and $\Omega_{M}$ from SNIa+CMB+SDSS joint analysis
done by Zhang and Wu (Fig. 2 of \cite{zhang-2007}).
The bright region at the center corresponds to the region
within the $1~\sigma$
confidence level contour.
The black dot denotes the best-fit point from the observations. The white dot represents our theoretical prediction
for the SM and the star  for the MSSM with $\Omega_M=0.27$ . The triangle denotes our prediction with $\Omega_M=0.29$ for the SM.
(Courtesy of F. Wu)
 \label{Fig1} }
\end{figure}
Remarkably, this theoretical value for $\omega^0_\Lambda$ is consistent with
 current observational data  from SN Ia, the cosmic microwave background (CMB), and
 the Sloan Digital Sky Survey (SDSS)
~\cite{movahed:083518,gong:043510,zhang:043524,1475-7516-2004-08-006,zhang-2007,wu2006,seljak-2006-0610}
(see Fig. 9 in ~\cite{movahed:083518} and Fig. 15 in ~\cite{wmap3} ) at the $95\%$ confidence level.
For example, the combination of  3-year WMAP  data and the Supernova Legacy
Survey data~\cite{wmap3} yields $\omega^0_\Lambda = -0.97^{+0.07}_{-0.09}$, which is in  agreement with our prediction,
although $\omega_\Lambda$ is assumed to be independent of time in that paper.
Very recently, Zhang and Wu~\cite{zhang-2007} perform a joint analysis of constraints
on $d$ with the latest observational data including  the gold sample of SN Ia,
the shift parameter of CMB  and the  baryon acoustic oscillation (BAO)
from the SDSS. This gives $d=0.91^{+0.26}_{-0.18}$,
 which  contains our value $d\simeq 0.95$ within the $1~\sigma$ region
(see  Fig. 1).
Thus, our model well explains observed properties of dark energy.
For the minimal supersymmetric standard model (MSSM), $N_{dof}=244$;
this value of $N_{dof}$  gives $d\simeq 1.36$ and  $\omega^0_\Lambda\simeq -0.75$,
which slightly violates the constraint $\omega^0_\Lambda<-0.76$ from SN Ia data~\cite{riess-1998-116}.

 This result indicates that, for our model, SM degrees of freedom is good for $N_{dof}$
and the Planck length scale is good for the UV cut-off $a$.
The reasons behind this might be as follows.
The origin of our entanglement energy
is different from the energy of an low energy effective theory considered
by Cohen et al's proposal \cite{PhysRevLett.82.4971}
which motivates the usual holographic dark energy models.
The entanglement energy is related to quantum information loss at the horizon
and to the vacuum quantum fluctuation in quantum gravity theory, which
 is usually believed as the origin of dark energy or the holographic principle.
Thus,
the natural UV cut-off of our model is the Planck length
 as in many related literatures~\cite{entenergy,Fursaev}.
What can we say about $N_{dof}$?
Considering the Planck scale UV cut-off, it is desirable to use
also the degree of freedoms at the Planck scale.
This value  depends on the model of the unification theory, which varies from $O(10^{2})$ to $O(10^3)$
and makes  the explicit value of $d$ vary approximately from  1 to  3.
However, SM is the only model that is  verified by various experiments so far.
Therefore, it is still plausible that
the degrees of freedom at the Planck scale could be similar to that of SM
and we can use SM degrees of freedom for $N_{dof}$.
Even in the case that  a larger unification theory (such as string theory)
is the true theory,
contributions from non-SM fields to vacuum fluctuation might be negligible due to symmetry breaking of those
sectors.
Although our theory still has some ambiguity to be resolved
in these parameters, it is interesting that our theory predicts the observed $d$ value
with the Planck scale and SM degrees of the freedom.

To obtain a more precise value   of $\omega^0_\Lambda$, it is essential to calculate
the exact value of $\beta=\beta_i$ for every field  $i$  in the SM or MSSM.
Then $\beta_i$ and the number of degrees of freedom of the $i$-th field, $N^i_{dof}$,
should satisfy the relation
$\sum_i \beta_i N_{dof}^i=4 \pi^2 d^2$,
derived by the same arguments as those leading to Eq. (\ref{d1}).
An interesting question here is whether the above equation gives $d=1$ for the SM.
Using Eq. (28) of Ref. \cite{li-2004-603}
one can also obtain the time dependency of the equation of state;
\beqa
\label{omega3}
\omega_{\Lambda} &=&\left (1+\frac{2\sqrt{\Omega_\Lambda}}{d}\right)
\left(-\frac{1}{3}
+z\frac{\sqrt{\Omega_\Lambda}(1-\Omega_\Lambda)}{6d}\right)\no
&=& -0.93+0.11 z
\eeqa
for SM,
where $z$ is the red shift parameter.

In general, holographic dark energy models  including ours
tacitly  assume
the presence of the accelerating expansion of the universe.
If not, the holographic dark energy
  could not be finite.
Since the accelerating universe is an observational fact, this assumption is plausible.
Alternatively, if we first assume finite $S_{Ent}$ at any finite time, then the accelerating universe is a natural consequence.
From Eq. (\ref{Sent}), finite $S_{Ent}$ implies a finite $R_h$.
 From Eq. (\ref{Rh}), it is easy to see that  the event horizon exists only when
$\int^\infty_t dt'/R(t')$
 converges, that is, the universe should accelerate ($R(t)\sim t^n, ~n>1$) as $t\rightarrow\infty$
 for  $R_h$ to be finite (unless the universe is oscillating)~\cite{Raychaudhuri}.
 The accelerating universe  satisfies
\beq
\label{friedmann}
\frac{\ddot{R}}{R}=-\frac{4\pi(\rho+3p)}{3}=n(n-1)t^{-2}>0,
\eeq
hence, $\rho+3p<0$, i.e., $\omega\equiv p/\rho<-1/3$.
Here the dot denotes a time derivative.
Thus, the finiteness of $S_{Ent}$ demands a finite $R_h$ and requires that the universe should accelerate
and  dark energy should dominate as $t\rightarrow \infty$.

There are already  many scenarios that explain the cosmic coincidence problem in the context of holographic dark energy.
For example, in ~\cite{RefWorks:5}
to solve the  coincidence problem,
an interaction between  dark matter~\cite{myhalo} and  dark energy
was introduced.
Li suggested  inflation at the GUT scale with
the minimal number of e-folds of expansion $N\simeq 60$
as a solution ~\cite{li-2004-603}.

In summary, we suggest a model in which   dark energy
is identified as the entanglement energy  of the universe.
This model could explain many observed properties of dark energy without
 modification of  gravity, exotic fields or particles.
 Using only  standard model fields, the holographic principle,
  and entanglement theory, our model predicts the
  equation of state and the constant $d$ of dark energy which are well consistent  with  observations.
Our analysis  also indicates that the holographic principle and the entanglement theory
can play a fundamental role
not only in
the physics of black holes
or string theory but also in cosmology~\cite{fischler-1998,bak-2000-17,freivogel:086003}.

\section*{acknowledgments}
Authors are thankful to  Changbom Park, Yun Soo Myung, and Yeong Gyun Kim for
helpful discussions. This work was partly supported by the IT R\&D program of MIC/IITA
[2005-Y-001-04 , Development of next generation security technology]
 (J.W. Lee), and by the Korea Research Foundation
Grant (MOEHRD, Basic Research Promotion
Fund) (KRF-2005-075-C00009;H.-C.K., and KRF-2006-312-C00095;
J.J. Lee) and by the Topical Research Program of APCTP
and the National e-Science Project of KISTI.



\end{document}